\documentclass[3p,times,procedia]{elsarticle}
\usepackage{nupha_ecrc}
\usepackage{amssymb}
\usepackage{graphicx}
\usepackage{color}
\usepackage{amsmath,amssymb}

\volume{00}

\firstpage{1}

\journalname{Nuclear Physics A}

\runauth{L.~Yan}

\jid{nupha}

\jnltitlelogo{Nuclear Physics A}

\usepackage{amssymb}

\usepackage[figuresright]{rotating}

\def\be{\begin{eqnarray}}
\def\ee{\end{eqnarray}}
\def\st{\begin{equation}}
\def\stp{\end{equation}}

\def\P{{\mathcal P}}

\def\bra{\langle}
\def\ket{\rangle}
\def\bbra{\langle\!\langle}
\def\kket{\rangle\!\rangle}

\def\Eq#1{Eq.~(\ref{#1})}

\def\Fig#1{Fig.~\ref{#1}}
\def\Sect#1{Section~\ref{#1}}
\def\Refs#1{Refs.~\cite{#1}}
\def\Ref#1{Ref.~\cite{#1}}

\begin{document}

\begin{frontmatter}



\dochead{XXVIth International Conference on Ultrarelativistic Nucleus-Nucleus Collisions\\ (Quark Matter 2017)}

\title{Hydrodynamic modeling of heavy-ion collisions}


\author{Li Yan}

\address{
Department of Physics,
McGill University\\
3600 rue University
Montreal, QC
Canada H3A 2T8
}

\begin{abstract}


This contribution presents a theoretical overview of hydrodynamic modelling of heavy-ion collisions, 
with highlights on some recent developments. In particular, the formulation of
anisotropic hydrodynamics, the role of 
hydrodynamic fluctuations, and the non-linear coupling of 
flow coefficients will be discussed.

\end{abstract}


\end{frontmatter}


\section{Introduction}
\label{sec:intro}

Heavy-ion collisions carried out at the Relativistic Heavy-Ion Collider 
and the Large Hadron Collider create a Quark-Gluon Plasma (QGP), a hot medium 
that
behaves like a perfect fluid. The fluidity of QGP is essential 
for understanding many of the
observed phenomena in high energy nucleus-nucleus collisions, 
in particular for soft probes such as long-range multi-particle correlations
whose dynamical properties are dominated by the bulk medium evolution. 

\begin{figure}[t]
\begin{center}
\includegraphics[width=0.4\textheight] {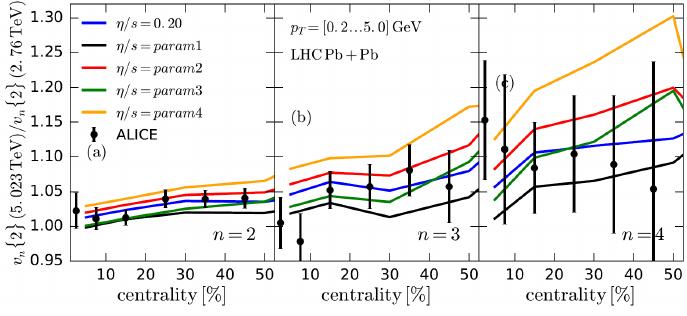}
\includegraphics[width=0.24\textheight] {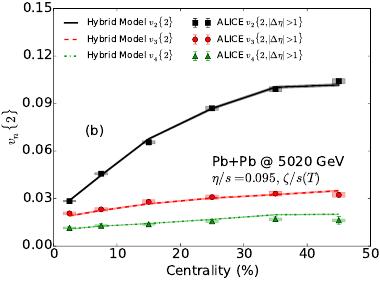}
\caption{
\label{fig:hydro_res}
Results of rms flow magnitude for harmonics $n=2$, 3 and 4 from 
two separate groups of hydro simulations with initial state generated by 
EKRT (left) and IP-Glasma (right), with respect to the most recent
measurements of Pb+Pb collisions at $\sqrt{s_{NN}}=5.02$ TeV.
Figures from \Ref{Niemi:2015voa} and \Ref{McDonald:2016vlt}.
}
\end{center}
\end{figure}

To quantify long-range
multi-particle correlations, the harmonic flow $V_n$ are often introduced.
Following a Fourier decomposition of the emitted 
single-particle spectrum in the azimuthal direction,
harmonic flow
\be
\label{eq:vn_def}
V_n= v_n e^{in \Psi_n} \equiv
\{ e^{in\phi_p}\}\,,
\quad n=1,2,3,\ldots
\ee 
characterizes momentum asymmetry in azimuth order-by-order. 
The curly braces in \Eq{eq:vn_def} notate average over 
the emitted particles of an individual collision event.
Note that the $V_n$ are defined as complex quantities in 
\Eq{eq:vn_def}, with magnitude and phase indicated by 
$v_n$ and $\Psi_n$, respectively.
For $n=2$ for instance, the so-called elliptic 
flow~\cite{Ollitrault:1992bk} $V_2$ describes asymmetric
emission of particles from directions in- and out-of reaction plane. 
In experiments,
extensive evidence of QGP fluidity has been obtained involving 
$V_n$ in various aspects in nucleus-nucleus collisions.
In addition to the simplest observables related to the rms average
of flow magnitude
$v_n\{2\}=\sqrt{\bbra |v_n|^2\kket}$ 
(double brackets $\bbra\ldots\kket$ 
stand for an event average), 
more precised details of harmonic flow have been
investigated in recent measurements.
These include fluctuations and 
correlations of the flow magnitude $v_n$ and correlations 
of phase $\Psi_n$, corresponding respectively 
to cumulant of harmonic 
flow $v_n\{m\}$ ($m=4$, 6, etc.), 
event-by-event flow distributions $P(v_n)$~\cite{Aad:2013xma}, 
symmetric cumulants $sc(n,m)$~\cite{ALICE:2016kpq} and 
event-plane correlators~\cite{Aad:2014fla}. 
Remarkably, an appropriate modeling of the medium 
evolution with hydrodynamics successfully describes all of the observed
flow signatures 
(see~\cite{Heinz:2013th,Luzum:2013yya} for recent reviews). 
\Fig{fig:hydro_res} presents 
results of 
hydro simulations for the most recent measurements of the rms
flow harmonics, for the 
Pb+Pb collisions at $\sqrt{s_{NN}}=5.02$ TeV, where
agreement between hydro and experiments is clearly seen.
One may refer to 
\Refs{Giacalone:2016afq,Zhu:2016puf,Gardim:2016nrr,Niemi:2015qia,Noronha-Hostler:2015uye}  
for hydro calculations of some other types of flow observables, regarding
recent measurements of the nucleus-nucleus collisions.


\section{Ingredients of hydrodynamic modeling}

Relativistic hydrodynamics plays a central role in 
the
modeling of heavy-ion collisions, in which
the evolution of QGP medium is described by solving an
appropriate equation of motion.
Hydrodynamics is an effective theory 
for the
low-energy (long wavelength) degrees of freedom of an evolving thermal
system, consisting of a set of conservation laws. The
conservation of energy-momentum, for instance, leads to hydrodynamic
equation of motion
\be
\label{eq:hydro_eom}
\partial_\mu T^{\mu\nu}=0\,,
\ee
where energy-momentum tensor $T^{\mu\nu}$ 
is constructed in terms of
hydrodynamic variables: Flow velocity $u^\mu$, 
energy density $e$, and pressure $\P$, 
\be
\label{eq:tmn}
T^{\mu\nu}= e u^\mu u^\nu - (\P+\Pi)\Delta^{\mu\nu} + 
\pi^{\mu\nu}\,.
\ee
$\Delta^{\mu\nu}=g^{\mu\nu}-u^\mu u^\nu$ is a projection tensor,
which helps to define the covariant form of spatial gradient,
$\nabla^\mu \equiv \Delta^{\mu\nu}\partial_\nu$.
The terms
$\pi^{\mu\nu}$ and $\Pi$ in \Eq{eq:tmn} are shear and bulk viscous
corrections to the stress tensor, respectively, containing a gradient expansion 
of hydrodynamic variables. 
To the first order in the 
expansion, the shear and bulk stress tensors are given by the
Navier-Stokes 
hydrodynamics, 
\be
\pi^{\mu\nu} = \eta \sigma^{\mu\nu}  + O(\nabla^2)\,,\qquad
\Pi=-\zeta \nabla\cdot u+O(\nabla^2)\,,
\ee
with coefficients shear viscosity $\eta$ and 
bulk viscosity $\zeta$ determined by the
underlying microscopic dynamics. For the case of QGP, it has been realized that
hydrodynamic description of the medium demands a small shear viscosity over entropy
ratio, $\eta/s\sim O(1/4\pi)$~\cite{Luzum:2008cw}, 
which is close to the theoretical expectation
for a strongly coupled system~\cite{Kovtun:2004de}. 
To avoid acausal modes in practical simulations,
second order viscous corrections are taken into account as well, which depend on
second order transport coefficients, 
$\tau_\pi$, $\tau_\Pi$, etc.
\Eq{eq:hydro_eom} is coupled and closed by an 
equation of state, for which 
results from lattice QCD are generally incorporated~\cite{Huovinen:2009yb}.

With respect to the system evolution in heavy-ion collisions,
hydrodynamics cannot be applied to 
the very early stages after 
collisions, during which the system is far from local equilibrium,
or very late stages when the system starts to decouple kinematically.
Despite 
a large local pressure anisotropy at the initial
stages of medium evolution (see \Sect{sec:ahydro} 
for more details), 
in hydro modeling 
an abrupt switch-on of hydro is taken into account around time
$\tau_0\sim O(1)$ fm/c, given an appropriate initial distribution
of energy (or entropy) density as input for the equation of motion,
\Eq{eq:hydro_eom}. In hydro modeling, initial density profile
can be realized by a variety of effectively models, such as 
MC-Glauber~\cite{Alver:2008aq},
MC-KLN~\cite{Drescher:2006ca}, EKRT~\cite{Niemi:2015qia} and 
IP-Glasma~\cite{Schenke:2012fw}. 
In these models, the energy (or entropy) deposition from collisions is realized through different prescriptions, but initial state fluctuations of nucleon-nucleon collisions are typically implemented on an event-by-event basis.
For instance, the IP-Glasma model
is inspired by the idea of gluon saturation~\cite{McLerran:1993ka}. 
Of particular
importance are the so-called initial state anisotropies, which are defined
through the energy density profile ($n\ge2$)~\cite{Teaney:2012ke},
\be
\label{eq:ecc}
\mathcal{E}_n=\varepsilon_n e^{in\Phi_n}\equiv
-\frac{\int dxdy\; e(x,y,\eta_s,\tau_0)(x+iy)^n}
{\int dxdy\; e(x,y,\eta_s,\tau_0)|x+iy|^n}\,.
\ee
$\mathcal{E}_n$ quantify geometric deformations of the initial density
profile, owing to the overlapped collision geometry
and fluctuations. Upon medium collective expansion, 
$\mathcal{E}_n$ provide the geometric origin of momentum anisotropies of 
the final state particle spectrum, namely, $V_n$. 
Regarding the calculations of flow harmonics, 
it is now widely acknowledged that
a large source of uncertainty of hydro modeling comes from
initial state. 
At late stages of the system evolution, 
hydrodynamics breaks down along with particle generation from the decoupled
medium (Cooper-Frye freeze-out~\cite{Cooper:1974mv}), 
which is further followed in the 
hydro modeling by hadron-cascade 
dominated evolution and resonance 
decay~\cite{Bass:1998ca,Bleicher:1999xi}.

\section{Recent developments in hydrodynamic modeling}
\label{}

\subsection{Anisotropic hydrodynamics}
\label{sec:ahydro}

\begin{figure}[t]
\begin{center}
\includegraphics[width=0.5\textheight] {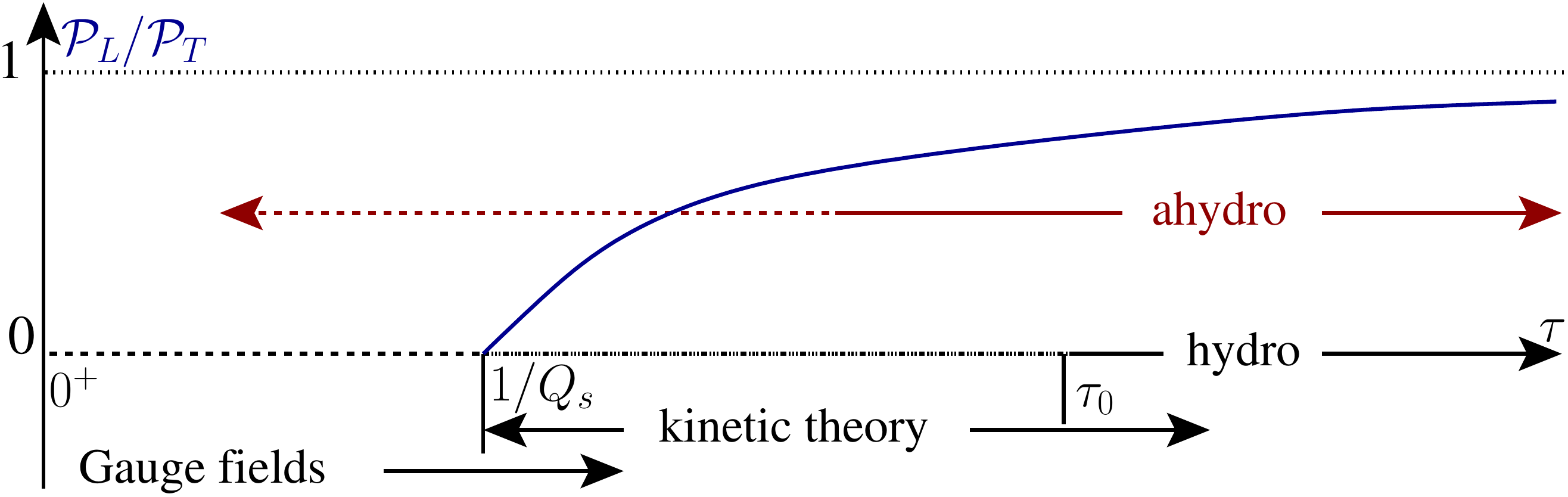}
\caption{ Pre-equilibrium evolution of a quark-gluon
system in heavy-ion collisions characterized in terms
of the pressure anisotropy $\P_L/\P_T$.
\label{fig:ic_evolve}
}
\end{center}
\end{figure}

Applicability of viscous hydrodynamics is constrained by the 
convergence of gradient expansion, which however is 
conceptually challenged by the issue of local pressure 
anisotropy. In the initial stages of heavy-ion collisions,
evolution towards local thermal equilibrium of a quark-gluon
system 
experiences strong longitudinal expansion along the beam
axis, which is reflected by the ratio of longitudinal
to transverse pressures, $\P_L/\P_T$.
These pressures are defined in kinematics as (assuming z-axis
as the beam axis)
\be
\P_L=\int \frac{d^3p}{(2\pi)^3p^0} p_z^2 f(t, \vec x,\vec p)
\,,\qquad
\P_T=\int \frac{d^3p}{2(2\pi)^3p^0} (p_x^2+p_y^2) f(t, \vec x,\vec p)\,,
\ee
so that for a locally thermalized system, when 
the phase-space distribution function gets isotropized, $\P_L/\P_T=1$. 
Since deviation from the thermalized distribution function in 
hydrodynamics is reflected in dissipation, 
pressure anisotropy is related to viscous corrections.\footnote{
For the case of Bjorken flow, it is known explicitly that
$\P_L-\P_T=-2\eta/\tau + O(1/\tau^2)$.
}
It is only with small viscous corrections,
namely when $\P_L/\P_T$ is close to unity, can one expect a good
convergence behavior of the gradient expansion. 
Illustrated in \Fig{fig:ic_evolve} is a qualitative picture of 
the time evolution of pressure
anisotropy in the initial stages of heavy-ion collisions, based on 
theoretical analyses of weakly coupled QCD systems (cf. \Ref{Keegan:2015avk}).
Pressure anisotropy is potentially large around 
$\tau_0\sim 1$ fm/c, 
although at which hydro starts to be applied in present modelings 
of heavy-ion collisions. 

\begin{figure}[t]
\begin{center}
\includegraphics[width=0.45\textwidth] {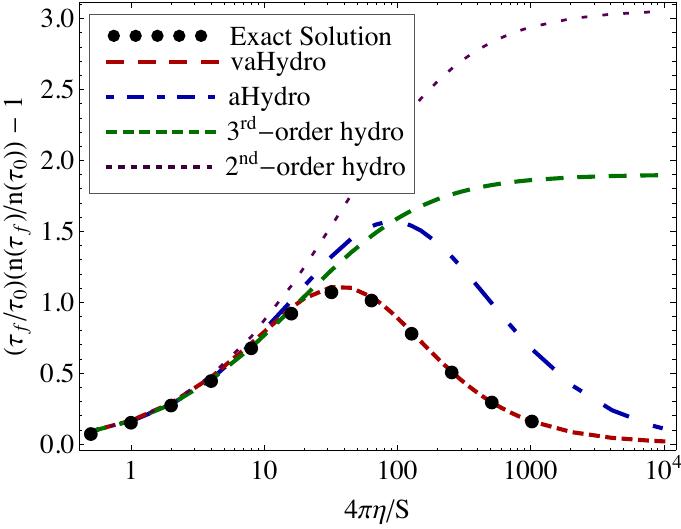}
\caption{
Convergence of ahydro and vahydro to the exact solution of 
Boltzmann equation with relaxation time approximation
for a 0+1D expanding system. Figure from
\Ref{Bazow:2013ifa}.
\label{fig:ahydro}
}
\end{center}
\end{figure}

Distinguished from a canonical formulation of viscous hydrodynamics 
(such as Chapman-Enskog expansion),
 in which the gradient expansion 
is realized with respect
to an isotropized distribution function, in the 
so-called anisotropic hydrodynamics 
(ahydro)~\cite{Martinez:2010sc,Florkowski:2010cf}, pressure anisotropy
is absorbed into an anisotropic background. Such a modification
greatly improves the applicability of hydrodynamics, especially
accounting for 
a large pressure anisotropy.
An example is shown in 
\Fig{fig:ahydro} for the calculation of number density, 
for a special case of
Bjorken flow (medium 
expansion dominated in the longitudinal direction) and 
relaxation time approximation. For a 
Bjorken flow and relaxation time approximation, the background
distribution in ahydro can be formulated as 
\be
\label{eq:ahydro_dis}
f(t,\vec x, \vec p) = f_{\mbox{\tiny iso}}
\left(\sqrt{\vec p^2+\xi(\tau)p_z^2}/\Lambda(\tau)\right)\,,
\ee
with momentum anisotropy captured by a new variable $\xi(\tau)$.
A gradient expansion around \Eq{eq:ahydro_dis} gives rise to the viscous 
anisotropic hydrodynamics (vahydro) in \Fig{fig:ahydro}. Comparing
to viscous hydrodynamics with 2nd or 3rd order viscous corrections, 
convergence of ahydro and vahydro to the exact solution is impressive.
Especially, one sees that vahydro applies to cases when
$4\pi\eta/s$ approaches to order of $10^3$, which is far beyond local
thermal equilibrium. One may refer to \Ref{Strickland:2014pga}
for a detailed review of ahydro.

Recent progress in ahydro 
formulation has been 
achieved in various 
aspects~\cite{Alqahtani:2017jwl,Nopoush:2015yga,Tinti:2015xra,Nopoush:2014qba,Florkowski:2014sfa}. 
For instance, it is realized that one can close ahydro equations of motion by an extra
equation matching 
the longitudinal pressure $\P_L$~\cite{Martinez:2017ibh,Molnar:2016gwq}. 

\subsection{Hydrodynamic fluctuations}

In addition to the initial state fluctuations which have quantum origins,
there are also thermal fluctuations during the whole evolution stages of
the quark-gluon medium.
In the absence of these thermal fluctuations in hydrodynamic simulations, 
hydro modeling is defective. Inclusion of thermal fluctuations in fluid
dynamics, namely, hydrodynamic fluctuations, has become a novel focus
in the heavy-ion community, considering especially its influence on the
studies of small colliding systems and the QCD critical point searching 
in the beam energy scan program. 

Hydrodynamic fluctuations can be formulated stochastically 
through the fluctuation-dissipation relations~\cite{Landau-sp2}. 
With respect to the energy-momentum
tensor $T^{\mu\nu}$, one is allowed to add a stochastic tensor $S^{\mu\nu}$,
\be
T^{\mu\nu}=e u^\mu u^\nu - (\P+\Pi)\Delta^{\mu\nu} +\pi^{\mu\nu}
+S^{\mu\nu}\,,
\ee
whose two-point auto-correlation is related to the corresponding dissipations.
For the Navier-Stokes hydrodynamics~\cite{Kapusta:2011gt},
\be
\label{eq:ss}
\bra S^{\mu\nu}(x)S^{\alpha\beta}(x')\ket
=2T\left[
\eta(\Delta^{\mu\alpha}\Delta^{\nu\beta}+\Delta^{\mu\beta}\Delta^{\nu\alpha})
+\left(\zeta-\frac{2}{3}\eta\right)\Delta^{\mu\nu}\Delta^{\alpha\beta}\right]
\delta^{(4)}(x-x')\,,
\ee
which clearly demonstrates that 
the strength 
of hydrodynamic fluctuations is
determined by shear viscosity and bulk viscosity. 
Generally, 
for a system with larger dissipation,
one would also expect a stronger effect of hydrodynamic fluctuations.

\begin{figure}[t]
\begin{center}
\includegraphics[width=0.95\textwidth] {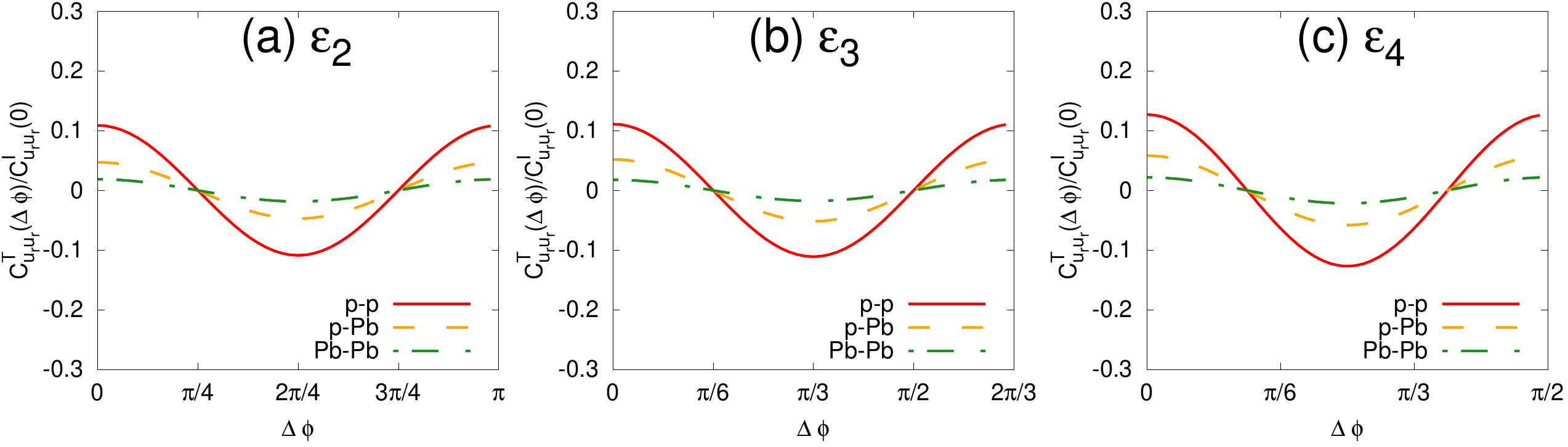}
\caption{
Additional increase of flow velocity correlations 
induced by hydrodynamic fluctuations in a 
1+1D expanding conformal system, with respect to
harmonic order $n=2$, 3 and 4, for the ultra-central
proton-proton, proton-lead and lead-lead systems.
Figures from \Ref{Yan:2015lfa}.
\label{fig:nhydro_res1}
}
\end{center}
\end{figure}

Despite the complexity of implementing hydrodynamic fluctuations in
a realistic
hydro modeling of heavy-ion collisions, there have been several attempts
towards an estimate of the effect of hydrodynamic fluctuations in
heavy-ion collisions. For a 0+1D expanding conformal system, inclusion of hydrodynamic
fluctuations can be 
reformulated to form a set of effective kinetic equations, which
couple to the background Bjorken flow~\cite{Akamatsu:2016llw}. 
It is then noticed that a fractional order,
$O(\nabla^{3/2})$,
in the gradient expansion arises from hydrodynamic fluctuations,
due to the non-linearities of the hydro equation of motion~\cite{Kovtun:2011np}. 
For a 1+1D expanding
conformal system, solution of the noisy viscous hydrodynamics can be 
achieved semi-analytically~\cite{Yan:2015lfa}, 
as a consequence of the analytically solvable
background Gubser flow~\cite{Gubser:2010ze,Gubser:2010ui}. 
In the context of heavy-ion collisions, it is 
first
interesting to notice that the two-point auto-correlation of 
hydrodynamic fluctuations, \Eq{eq:ss}, boils down to a form which
inversely
depends on the event-averaged multiplicity. Such a statement
supports application of hydro modeling to small colliding systems,
provide that multiplicity productions in these collisions are 
sufficiently high. Taking into account simultaneously contributions
from initial state fluctuations, one can further explore the effect
of hydrodynamic fluctuations on the flow harmonics. In \Fig{fig:nhydro_res1},
the additional increase of flow velocity anisotropies due to
hydrodynamic fluctuations is plotted
for the ultra-central proton-proton, proton-lead and lead-lead 
systems. Although the overall increase is not significant (smaller in
lead-lead, larger in proton-proton), 
a clear trend of enhancement from the second order harmonics
to the fourth order harmonics exhibits.

\begin{figure}[t]
\begin{center}
\includegraphics[width=0.35\textwidth] {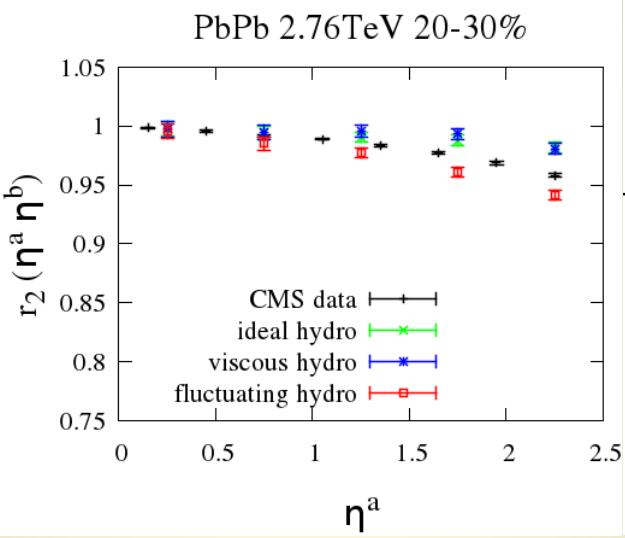}
\includegraphics[width=0.35\textwidth] {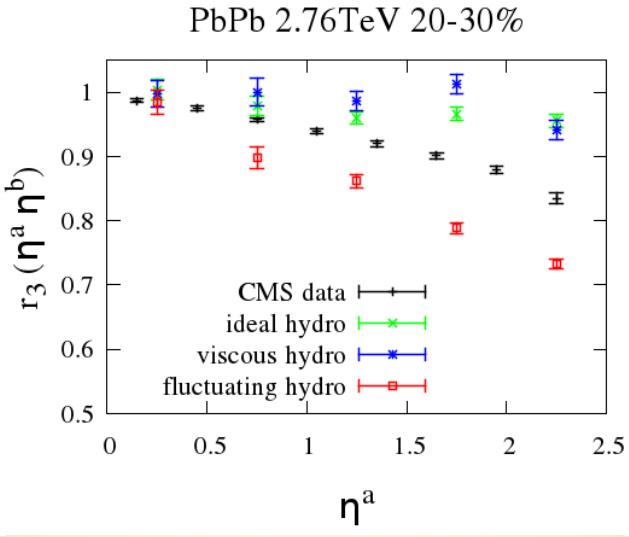}
\caption{
Factorization breaking of the second and the third order harmonics,
calculated from event-by-event hydrodynamics with also hydrodynamic
fluctuations. Figures from \Ref{Sakai:2017}.
\label{fig:nhydro_res2}
}
\end{center}
\end{figure}

The statement that hydrodynamic fluctuations are more significant 
for higher order harmonics gets also examined in realistic numerical
simulations~\cite{Sakai:2017}. \Fig{fig:nhydro_res2}
presents the results of factorization ratio of flow harmonics of
order 2 and 3, as a function of pseudo-rapidity, 
from one of the numerical hydro simulations containing hydrodynamic fluctuations.
Breaking of the factorization reflects event-by-event fluctuations
in the hydro modeling.
In addition to the breaking owing to initial
state fluctuations, extra contributions are induced from hydrodynamic
fluctuations, which are stronger in $v_3$ than $v_2$.

\subsection{Nonlinear mode couplings in harmonic flow}

Hydro modeling of the medium evolution 
in heavy-ion collisions is remarkably successful in
calculating harmonic flow. A flow paradigm is established 
accordingly based on event-by-event simulations for 
nucleus-nucleus collisions,
from which
relations among the harmonic flow $V_n$, initial state geometric
anisotropies $\mathcal{E}_n$, and the 
dynamical properties of the fluid medium, 
is proposed. 
Driven by the recent progress
towards a precision measurement of flow harmonics, 
this flow paradigm needs to be examined 
at a more detailed level in the application of hydrodynamics.
In particular, by doing so, one would expect to 
reduce uncertainties on the extracted transport 
coefficients of QGP. 

Initial anisotropies in \Eq{eq:ecc} are small quantities 
bounded by one, by definition. A linear eccentricity 
scaling relation can thereby be
introduced between $V_n$ and $\mathcal{E}_n$, by cutting 
the expansion over $\mathcal{E}_n$ at the linear order. Indeed,
hydrodynamic simulations have verified that, to a good 
approximation, in nucleus-nucleus 
collisions~\cite{Niemi:2012aj,Gardim:2011xv},
\be
\label{eq:linear}
V_n=\kappa_n(\eta,\zeta) \mathcal{E}_n\,,\quad n=2,3,
\ee
when collision centrality percentile is not very large.
Note that the information of the initial state geometry is
contained separately in $\mathcal{E}_n$, the coefficient
$\kappa_n$ which we refer to as the linear response coefficient
depends only on medium dynamical properties. Given this linear
relation, one is allowed to relate, for instance, the event-by-event 
fluctuations of flow $P(v_n)$,
directly to that of initial 
anisotropies 
$P(\varepsilon_n)$~\cite{Giacalone:2016eyu,Yan:2014afa,Castle:2017}. 

\begin{figure}
\begin{center}
\includegraphics[width=0.5\textwidth] {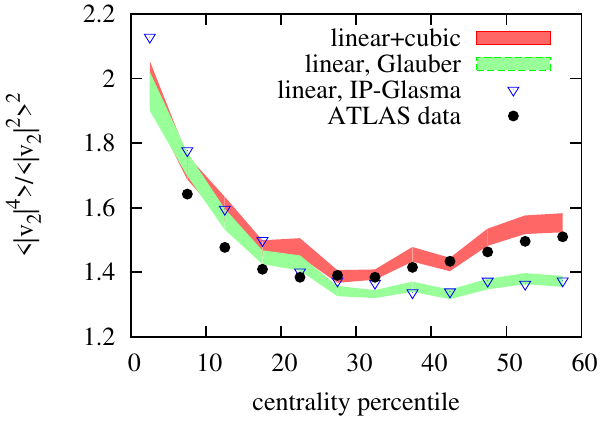}
\caption{
Ratio of fourth order moment to the square of the rms value
of $v_2$ magnitude. Experimental results are compared  
with event-by-event hydrodynamic simulations, and initial
models where cubic response is absent. Figure
from \Ref{Noronha-Hostler:2015dbi}.
\label{fig:v2_ratio}
}
\end{center}
\end{figure}

There are two cases, however, in which one has 
to consider beyond the 
linear eccentricity scaling relation in hydro modelings. 
The first involves lower order flow harmonics ($n=2$, 3) in 
nucleus-nucleus
collisions of large centralities. In these collision events,
geometry of the overlapped region is so deformed that 
$\varepsilon_2$ cannot be treated as small. Constrained by
rotational symmetry and analyticity of the relation, 
the correction to the linear eccentricity scaling relation
of $V_n$ is of cubic order,
\be
\label{eq:cubic}
V_n=\kappa_n \mathcal{E}_n+\kappa_n'|\mathcal{E}_2|^2\mathcal{E}_n\,,\qquad
n=2,3\,,
\ee
where $\kappa_n'$ is the cubic order response coefficient.
Event-by-event hydro simulations have been carried out which
found
supports to the relation in \Eq{eq:cubic}~\cite{Noronha-Hostler:2015dbi}. 
The role of cubic
response is not obvious in the rms magnitudes of elliptic flow 
$V_2$. However, in the more differential analysis of the event-by-event
$v_2$ distributions, it is noticed that cubic response is
crucial in explaining the observed cumulant ratios.  
In \Fig{fig:v2_ratio}, the ratio between the 
fourth order moment and square of the rms value of $v_2$ is plotted as 
a function of centrality percentile, for the Pb+Pb 
collisions at the LHC energy. Linear eccentricity scaling 
implies that the ratio is identical in $v_2$ and 
$\varepsilon_2$. Comparing to the results calculated 
from the initial state modelings, the apparent difference observed when
centrality becomes larger than $30\%$ indicates the significance
of cubic response.

In the other case concerning higher order flow harmonics ($n\ge 4$), 
non-linearities are more essential.
It is the consequence that 
the linear part of higher order flow
get
strongly suppressed by viscous 
corrections~\cite{Teaney:2012ke}.
Following the similar strategy
mentioned before, one writes flow harmonics in an expansion
of the couplings of $\mathcal{E}_2$ and $\mathcal{E}_3$, 
with respect to rotational symmetry in the azimuthal direction.
Further re-expressing $\mathcal{E}_2$ by $V_2$, and $\mathcal{E}_3$
by $V_3$, one has 
for $V_4$ and $V_5$ a separation between the linear part and the
nonlinear couplings~\cite{Yan:2015jma},
\be
\label{eq:nlinear}
V_4=V_4^L + \chi_{422} V_2^2 \,,\qquad
V_5=V_5^L + \chi_{523} V_2 V_3\,,
\ee
where we have defined nonlinear flow response coefficients
$\chi_{422}$ and $\chi_{523}$. Similar relations can be
obtained for $V_6$ and $V_7$, where multiple nonlinear
coupling terms and nonlinear flow response coefficients present.
One notices that $\chi_{422}$ and 
$\chi_{523}$ are measurables in experiments which 
rely only on medium properties,
\be
\label{eq:chi}
\chi_{422}=\frac{\bbra V_4 {(V_2^*)}^2\kket}{\bbra |V_2|^4\kket}\,,
\qquad
\chi_{523}=\frac{\bbra V_5 (V_2^*V_3^*)\kket}{\bbra |V_2V_3|^2 \kket}\,,
\ee
with the assumption that
the linear part and the nonlinear part of $V_4$ and $V_5$ 
are uncorrelated~\cite{Yan:2015jma,Qian:2016fpi}. 
These new observables have been measured 
in experiments at the LHC 
energy~\cite{Zhou:2017,Tuo:2017}, in comparison with
results from recent hydro simulations.
\Eq{eq:nlinear} can also be applied to more complicated flow 
signatures involving higher order flow. For instance, it is
realized that the symmetric cumulant, $sc(n,m)$ which measures
mixings of flow magnitudes, can be factorized into event-by-event
flow
fluctuations and event-plane correlations~\cite{Giacalone:2016afq}.

\section{Summary}

The success of hydrodynamic modeling of heavy-ion collisions
is perhaps 
the most convincing evidence of the QGP fluidity
in nucleus-nucleus collisions. In recent measurements
at different colliding energies at RHIC
and the LHC, soft probes of various types 
associated with the evolving medium were explored 
at an unprecedented level of resolution,
which further pushes the limits of hydro modeling.
Anisotropic hydrodynamics
is developed, which abandons local pressure
isotropization at the initial stages of the medium evolution.
Effect of hydrodynamic fluctuations has  been analyzed
as well. At a qualitative level, a new term of the
order of $O(\nabla^{3/2})$ is discovered in the gradient expansion
due to hydrodynamic fluctuations. Quantitative predictions are 
achieved regarding two-particle correlations in heavy-ion collisions, 
with more realistic simulations
containing hydrodynamic fluctuations carried out.
In addition to these theoretical developments, a flow 
paradigm which systematically analyzes the nonlinear couplings 
of harmonic flow is established. 
The fluid dynamical modelling of nuclear collisions continues to garner empirical success. The developments discussed herein should take this framework to its next level of sophistication.

\section*{Acknowledgements}
LY thanks the organizers of Quark Matter 2017 for the invitation to 
present an overview on hydrodynamic modeling in heavy-ion collisions.
LY also thanks C. Gale and S. Jeon for carefully reading the manuscript.
This work is supported in part by the Natural Sciences and Engineering
Research Council of Canada.





\bibliographystyle{iopart-num}
\bibliography{refs-qm17}







\end{document}